

  \MAINTITLE={The absorbers towards Q0836+113\FOOTNOTE{Based
on observations made with the Nordic Optical Telescope}
}
%
%

  \AUTHOR={N. Ben\' \i tez@1@2, E. Mart\' \i nez-Gonz\'alez@1, 
J.L. Sanz@1, A. Aguirre@3 and M. Alises@3 }

    \OFFPRINTS={N. Ben\'\i tez }

  \INSTITUTE={@1 IFCA, CSIC-UC,
                 Fac. de Ciencias, Avda. Los Castros s/n, 
                 39005 Santander, Spain
              @2 Dpto. de F\' \i sica Moderna, UC,
                 Fac. de Ciencias, Avda. Los Castros s/n, 
                 39005 Santander, Spain
              @3 Centro Astron\'omico Hispano-Alem\'an,
                 Calar Alto, Almer\' \i a, Spain
}
  \DATE={Received June 2, 1996; accepted October 2, 1996}
  \ABSTRACT={We have performed $RIJHK_S$ imaging of the field around the 
$z=2.67$ quasar Q0836+113, which presents several metal line and a 
damped Ly$\alpha$ absorption systems in its spectrum. The images reveal 
the existence of a red $K_S=18.9$ object $\approx 
11 \arcsec$ from the quasar. On the basis of the empirical 
relationships between absorption radius and luminosity we conclude
that this object may be the CIV absorber at $z=1.82$. This could be 
the first detection of a high redshift galaxy causing 
high-ionisation absorption.
After carefully subtracting the QSO, we do not detect, up to 
a $3\sigma$ limiting magnitude for extended objects of $K_S=20.8$,
the damped Ly$\alpha$ absorber apparently detected
as a Ly$\alpha$ emitter at $z=2.47$. This imposes 
an upper limit on its H$\alpha$ emission comparable to the results 
obtained  spectroscopically by Hu et al. (1993). 
It is also suggested, that object ``SW'' from  Wolfe et al. (1992) 
could be the galaxy responsible for the claimed MgII absorption at $z=0.37$.
}
\KEYWORDS={Galaxies:evolution -- Galaxies:halos -- $~~~~~~~~~~$
quasars:individual:Q0836+113 -- quasars:absorption lines }
\THESAURUS={03}
\maketitle
\MAINTITLERUNNINGHEAD{The absorbers towards Q0836+113}
\AUTHORRUNNINGHEAD{N. Ben\'\i tez et al.}
\titlea{Introduction}

The absorption line systems found in the spectra of distant quasars have 
proved to be very useful in order to find and study 'normal' high 
redshift galaxies, not flagged by the prominent emission of an AGN.
Several authors have followed this strategy, performing
broad band imaging and follow up spectroscopy of objects in quasar fields 
or with narrow band filters. The success of
absorber identifications up to $z \la 1$ has been very high: 
MgII systems are associated to galaxies which are 
identified in practically all the cases (Bergeron \& Boisse 1991, 
Steidel et al. 1994, 
Steidel 1995) and Ly$\alpha$-clouds are also identified 
with galaxies in a high fraction ($\simeq 50\%$) by Lanzetta et al. (1995)
. 

In this paper we report the results of deep near-IR and optical imaging 
of the field of the $z=2.67$ quasar Q0836+113, which presents several
low and high ionisation absorption systems and a damped Ly$\alpha$ 
system in its spectrum.

\titlea{Observations and photometry}

Optical $R$ and $I$ observations were carried out at the 2.6m Nordic Optical 
Telescope (NOT) at La Palma on the nights of 1995 
March 3, 4 and 5. The detector was the Thomson 
$1024^2$ IAC camera with a pixel scale of $0.14 \arcsec$/pixel. 
The field was centered on the quasar and we took 
3 frames of 2700 seconds each in $R$ and 5 frames of 1800 seconds in $I$.
The images were reduced with dome and sky flats using IRAF tasks 
and calibrated with standard stars. 
We estimate the zero-point calibration errors to be $\pm 0.06$ in $R$ 
and $\pm 0.08$ in $I$. The FWHM of the final co-added images is 
$1.8\arcsec$ and $0.9\arcsec$ for $R$ and $I$ respectively. 

   The near IR observations were performed at the 3.5m Telescope at Calar Alto
with the camera MAGIC on the nights of 1995 March 17, 19 and 21. MAGIC uses 
a $256\times256$ NICMOS3 HgCdTe detector array with $0.33 \arcsec$/pixel. 
Due to a telescope pointing error the field is not centered on the 
QSO, but displaced to the SW. 18 frames of 100 seconds and 9 of 290 
seconds were taken on $H$ and $J$ respectively and 81 frames of 60 seconds in 
the $K_S$ band. 
The images have been reduced in the standard way for near IR observations.
After subtracting dark frames, a flat field for 
each frame was formed by median combining the closest frames with a 
sigma-clipping algorithm. 
The final, averaged $K_S$ image, has a $0.8\arcsec$ 
seeing, and the $H$ and $J$ images have seeings of $1.5\arcsec$. The frames
were calibrated with standard stars from Elias et al. (1982).
The zero-point uncertainties are 3\% for $K_S$ 
and 10\% for $J$ and $H$. 

We have selected as possible candidates for the absorbers 
the objects detected in the $K_S$ image, which is the deepest and 
the one with best seeing.  
We used the package PISA (Draper \& Eaton 1992
) with a $3\sigma$ detection limit 
and carefully checked all detections by eye, removing spurious objects.
on the outer parts of the image, which are less exposed.
Afterwards we performed aperture photometry centered on these detections 
in the $R$, $I$ and $K_S$ bands using three apertures for each object: 
1.5$\sigma_{psf}$, 3$\sigma_{psf}$ and 6$\sigma_{psf}$, where 
$\sigma_{psf}$ is the $\sigma$ of the point spread function for each image.
\begfig 8.8cm
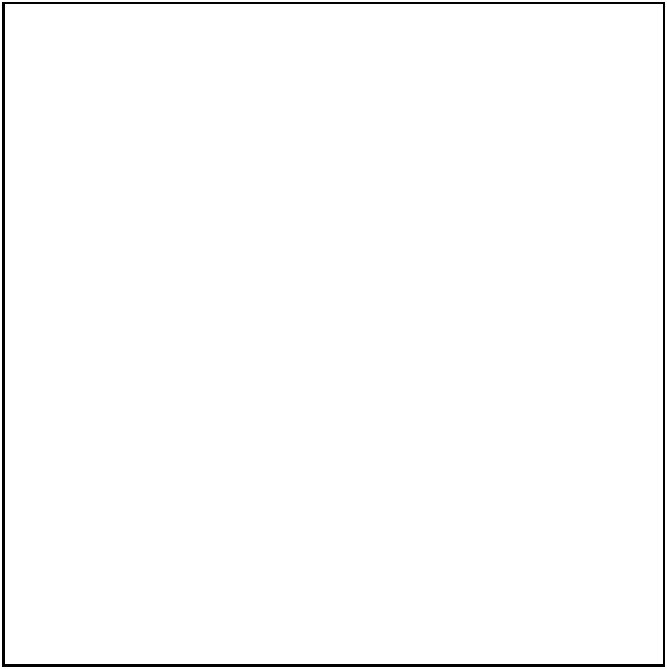
\figure{1}{$I - K_S$ vs. $K_S$ plot for all the objects detected in our field 
at a higher than $3\sigma$ level. The error bars include both the intrinsic 
photometric error and the rms of the corrections.}
\endfig
We classified the objects into two classes: stellar and extended. Extended
objects have on average, after quadratically subtracting the seeing, an 
angular extension of $\approx 1\arcsec$. For each object we took the magnitude 
given by the largest aperture for which the intrinsic photometric error 
was less than $0.3$ mag and, if needed, corrected to a 6$\sigma_{psf}$ 
aperture with the average of the corrections for the brighter objects of 
each class. 
The 6$\sigma_{psf}$ radii correspond to aperture diameters of 
$8.8\arcsec$ in $R$, $4.6 \arcsec$ in $I$ and $4 \arcsec$ in $K_S$.
In Fig. 1 we have plotted $I-K_S$ colors as a
function of $K_S$ for the objects detected in our $K_S$ frame. 
The $J$ and $H$ images are not so deep as the other ones and only overlap
with them in the central part containing the QSO. The results of the 
$RIJHK_S$ photometry of the QSO and the closest objects are presented in 
Table 1. 
\begtabfullwid
\tabcap{1}{Coordinates, distance from Q0836+113 and $RIJHK_S$ magnitudes}
\halign{#\hfil&&\quad#\hfil\cr
\noalign{\hrule\medskip}
Object &
X('')&
Y('')&
Radius('') & 
$R$ & 
$I$ & 
$J$ & 
$H$ & 
$K_S$ \cr
\noalign{\medskip\hrule\medskip}
QSO  &    0.   & 0 & 0 & $19.426\pm 0.002$ &$19.105\pm 0.007$ &
$18.69\pm 0.05$ &$18.17\pm 0.15$   &$17.59\pm 0.02$ \cr
C      &  -0.83 & 3.41 & 3.51  &$22.37\pm  0.16$  &$22.60\pm  0.15$  &
$>21.4 $   &$>20.2 $    &$20.11\pm 0.22$ \cr
SW     &  4.26 & -5.19 & 6.71  &$23.39\pm  0.19$  &$23.10\pm  0.21$  &$>21.4
$ &$>20 $  &$20.35\pm 0.19 $ \cr
S      & 0.72 & -10.64 & 10.66  &$23.63\pm  0.23$  &$22.73\pm  0.14$  
&$21.10\pm 0.20$   &$19.74\pm 0.20$    &$18.91\pm 0.08$ \cr
\noalign{\medskip\hrule\medskip}}\endtab
We have also plotted in Fig. 2 the spectral distributions,
inferred from broad band colours, of the QSO and object S. 
In order to check the photometry we fitted a $F_\nu \propto \nu^\alpha$ 
distribution to the QSO spectra and obtained $\alpha=-0.21$, close to 
the median slope of the QSOs observed by Neugebauer et al. 1987. The fit is 
overplotted in Fig. 2. 
\begfig 8.8 cm
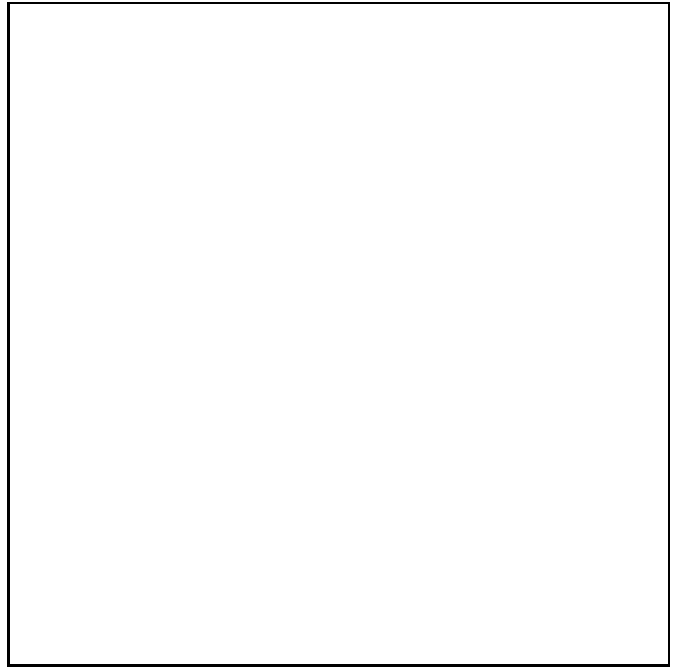
\figure{2}{Spectral distributions, inferred from broad band colours, of 
the QSO (above) and object S (below). A fit to a $F_\nu \propto \nu^\alpha$ 
distribution for the QSO and to a model galaxy at z=1.82 corresponding to
object S are shown (see details in the main text).}
\endfig
\titlea{Results and Discussion}
     
Turnshek et al. (1989) reported the presence of four absorption line 
systems in the spectra of Q0836+113: two MgII systems, one at $z=0.37$ and the 
other at $z=0.79$; a CIV system at $z=1.82$ which has no low-ionisation 
associated absorption and a damped Ly$\alpha$ (DLA) system at $z=2.47$. 
It should be pointed out that the MgII at $z=0.37$ lies in the 
Ly$\alpha$ forest and thus there is a chance of misidentification.
We have examined the best available data (Hunstead et al. 1990), 
kindly provided by M. Pettini, and could not confirm nor 
rule out definitively the reality of this absorption.

Hunstead et al. (1990) claimed the detection of a Ly$\alpha$ emission spike 
at the bottom of the DLA absorption line. This result generated 
considerable controversy: neither Wolfe et al. (1992) nor Lowenthal et al. 
(1995) have found traces of Ly$\alpha$ emission from the DLA absorber 
in their spectroscopic observations, although Pettini et al. (1995) 
apparently have confirmed the detection. Besides, Wolfe et al. (1992) 
reported the discovery of a faint extended Ly$\alpha$ emission a 
few arcsec to the NE of the QSO which has not been confirmed, 
neither spectroscopically nor through direct imaging by Lowenthal et al.(1995).
\begfig 8.8cm
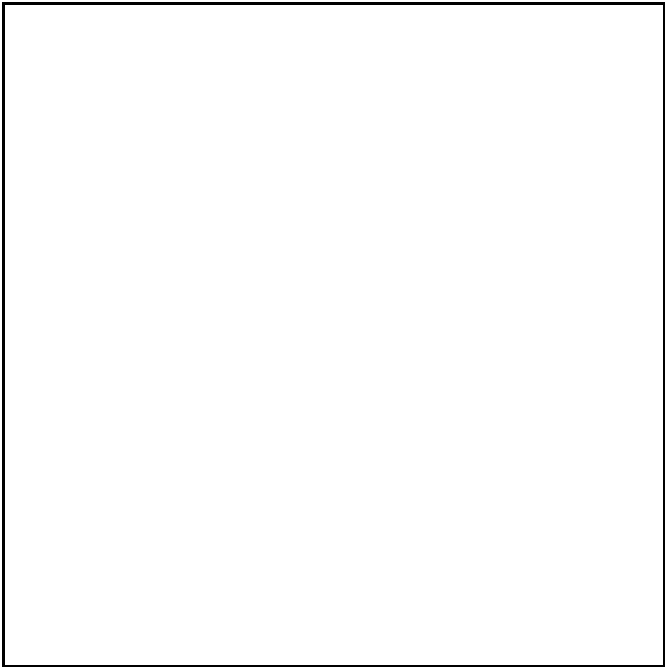
\figure{3a}{$I$-band image of a $\approx 36\times 36$ arcsec$^2$
field centered on Q0836+113. The image has been convolved with the seeing.
North is up and East to the left. The pixels have a scale of 
0.33$\arcsec$/pixel. }
\endfig
In Figs. 3a and 3b we present $I$ 
and $K_S$ images of a $\approx 36\times36$ arcsec$^2$ field 
centered on Q0836+113. 
\begfig 8.8cm
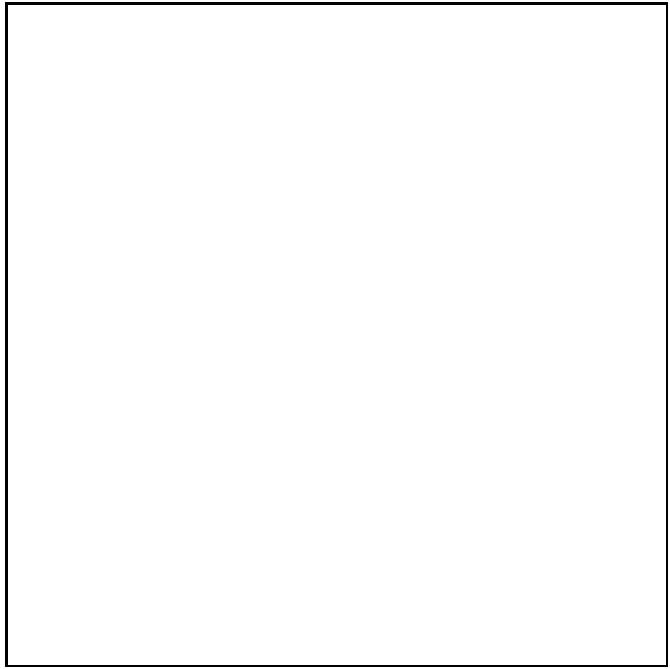
\figure{3b}{The same as Fig 3a, but in the $K_S$ band.}
\endfig
In order to check for the existence of possible objects which 
could be hidden by the quasar, we scaled a nearby, bright but 
non saturated star, and subtracted it from the quasar in the $K_S$ frame. 
The result can be seen in Fig. 3c. No object is detected 
at the position indicated by Wolfe et al. (1992) for the Ly$\alpha$ 
emitting object 
up to a $3\sigma$ detection limit (for extended objects in an
4\arcsec aperture) of $K_S=20.8$. This 
apparent magnitude limit cannot be translated to rest-frame luminosity
without big uncertainties, but for comparison we would have marginally 
detected in our image object B2 from Francis et al. (1996) which 
has $z=2.38$ and a Ly$\alpha$ flux similar to that reported by Wolfe 
et al. (1992) for their detection. 
\begfig 8.8cm
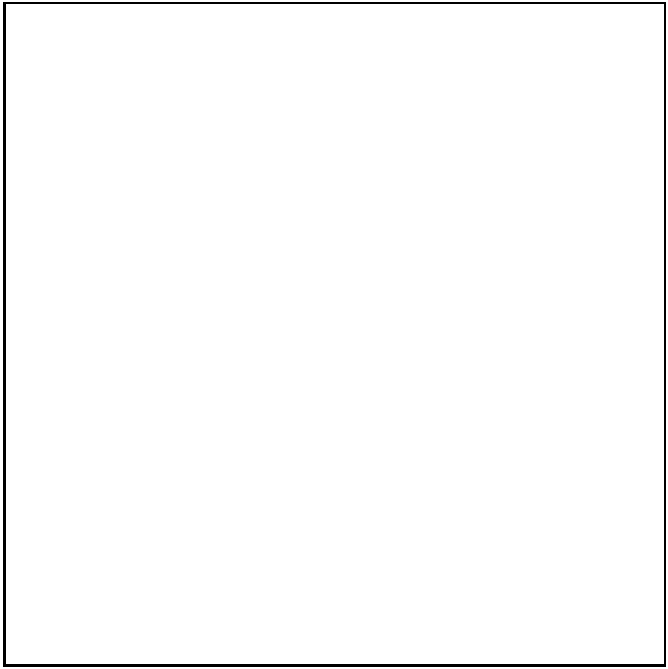
\figure{3c}{The same as Fig 3b, but after subtracting the QSO using the
star at the lower left corner of the field.}
\endfig
It is also noteworthy that we can impose upper limits
on the possible redshifted H$\alpha$ emission from the DLA 
($\lambda\approx 2.27\mu$m) as it is contained within the $K_S$ band 
($2.00-2.32\mu$m). Our 3$\sigma$ detection is $K_S=21.75$ arcsec$^{-2}$, 
corresponding to a 3$\sigma$ detection flux of 
$f_{K_S}\approx 2.7\times10^{-16}$ ergs cm$^2$ s$^{-1}$ arcsec$^{-2}$, 
which compares well with the 2$\sigma$ upper limit of 
$\leq 2.4\times10^{-16}$ ergs cm$^2$ s$^{-1}$ obtained by Hu et al. (1993) 
spectroscopically for the H$\alpha$ emission of this quasar. 

In order to find out how the objects detected in the frame 
are related to the QSO absorption systems,
we shall use the results of Steidel and collaborators 
(Steidel 1995, Steidel et al. 1994) who have 
identified the galaxies producing $ z \leq 1$ MgII absorption 
in a sample of quasars. They have established that if a galaxy is brighter 
than $M_K=-22$ (H$_o$=50 km s$^{-1}$ Mpc$^{-1}$, $q_o$=0.05) and falls 
within the distance 
$$R_{abs}(L_K)=38h^{-1}(L_K/L^*_K)^{0.15}$$
from a QSO sight line, it will 
produce detectable MgII absorption, independently of its  
morphological type. 
\begfig 8.8 cm
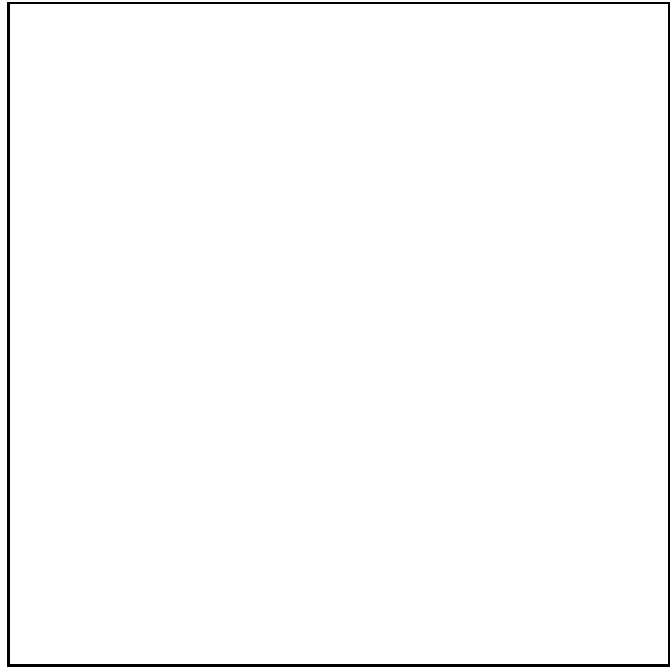
\figure{4}{Steidel et al. (1994) relationship shown for several redshifts.
The filled squared points represent the distance from the QSO and the apparent
$K$ magnitude of the objects detected within a $ 20\arcsec $ radius from 
the QSO. The thick lines show the low-ionisation absorption radius as a 
function of the $K$ apparent magnitude for several redshifts. The 
short-dashed line corresponds to $z=0.368$, the continuous line to
$z=0.788$ and the dotted line to $z=1.82$. The thin dotted line 
corresponds to the maximal radius of high-ionisation absorption at $z=1.82$
 (see text)}
\endfig
In Fig. 4 we have drawn $R_{abs}$ as a function of $K$ apparent 
magnitude for the redshifts $z=0.368$, $z=0.788$ and $z=1.822$. 
We have used the cosmological parameters 
($q_o$=0.05, H$_o$=50 km s$^{-1}$ Mpc$^{-1}$ ) and 
the K-corrections corresponding to a type Sb-Sc galaxy.
We have also plotted the objects which are within a radius of 
20$\arcsec$ from the QSO. Below we discuss the individual 
results regarding the most interesting objects in the field.

\titleb{Object C}
Wolfe et al. (1992) reported the detection of object C nearly superposed 
with the claimed NE Ly$\alpha$ emission. This galaxy
has been identified as the MgII absorber at $z=0.788$ (Lowenthal et al. 1995). 
In Fig. 3 object C can be clearly distinguished, $3.5\arcsec$ from the QSO
and at a PA of $\approx 7.5\deg$. We resolve it partially and find it 
to be quite elongated, with an ellipticity $e=0.28$. It has a bright 
nucleus closer to the QSO, and a very faint tail extending NW can be 
guessed. As could be expected, object C is well within the line 
corresponding to $z=0.788$ MgII absorption in Fig. 4.

\titleb{Object SW}
Wolfe and collaborators also detected the point-like
object SW, at $\approx 7 \arcsec$ in the SW direction from the 
QSO which we can also distinguish in Fig 3. 
In Fig. 4 we see that, excluding object C, only the position of object SW 
is consistent, within the error bars, with being the MgII absorber at 
$z=0.368$. This tentative identification should be confirmed 
spectroscopically, since this galaxy has not been detected in Fabry-Perot 
maps tuned to [OII]3727 at $z=0.37$ (Caulet 1991) and
Wolfe et al. (1992) report that it apparently 
presents a relative excess of emission in a narrow band filter tuned to
Ly$\alpha$ at the redshift of the DLA system. 
Besides, the galaxy itself is rather faint for $z=0.368$, 
as its $K=20.35\pm0.2$ corresponds to $L_K \approx 0.03L^*_K$, 
which is below the luminosity at which Steidel and coworkers suggest 
that the luminosity/cross-section 
scaling relation must break down, $L_K \approx 0.05L^*_K$. 
On the other hand, our detection limit of $K_S=20.8$
corresponds to an absolute magnitude of  $M_K\approx -20.6$, and
therefore the galaxy responsible for the absorption should be in 
our frame unless it is even fainter, $L_K \approx 0.02L^*_K$.
In any case, we must keep in mind that the existence of the MgII 
absorption at $z=0.37$ is still uncertain, and thus it could be
that there is no absorber galaxy at all.

\titlec{Object S}
We can clearly see in our frames an object 
$10.6 \arcsec$ to the south of the QSO which was not mentioned in
Wolfe et al. (1992) and Lowenthal et al. (1995), and which we call object S. 
It has a $K_S$ magnitude of $18.9\pm 0.06$ and is detected in all 
our broad-band images. It is red ($R-K_S$=4.7), partially resolved 
(FWHM$\approx 1\arcsec$) and slightly elongated in the NE direction. 
From Fig. 4 it is clear that object S is not close enough 
to the QSO to be a MgII absorber at any of the considered redshifts. 
As we explain below, the analysis of the optical-near IR colours 
of object S agree with this result and makes very unlikely that this 
object is a companion to these MgII absorbers. 
In Dickinson (1995) and Arag\'on-Salamanca et al. (1993)
we can find the $I-K$, $K$ color-magnitude relationship for Cl 1603+4313, 
a cluster of galaxies at $z=0.895$. 
Object S, with $I - K = 3.8$ is $\approx $0.6 magnitudes 
redder than the red envelope of this cluster. Assuming that this color 
is not caused by dust-absorption, it indicates that object S has $z>1$. 

Could object S be the CIV absorber at z=1.82?.
For illustration, 
we have plotted over the observed spectral distribution 
in Fig. 2b a model spectrum at $z=1.82$ obtained with the 
Bruzual \& Charlot (1993) code.
The model galaxy is formed by a burst of star formation involving
most of the galaxy mass at $z=7$ and undergoes another burst of star 
formation at $z=1.82$. 
The squares represent the empirical fluxes through each of the filters 
with their errors, the continuous line is
the model spectrum and the triangles are the fluxes that we would measure
after convolving this spectrum with the response of the filters. The 
galaxy spectrum is normalized to coincide with our data in the $K-$band.
We see that the colors of object S are consistent with $z=1.82$. 
At this redshift, object S would be rather luminous 
$L_K \approx 4.5L^*_K$ and at 
a distance of $\approx 127$ kpc (H$_o$=50 km s$^{-1}$ Mpc$^{-1}$, $q_o$=0.05) 
from the quasar position. 

In contrast with the growing number of galaxies known to be responsible 
for low-ionisation absorptions, so far there has not been any single 
identification of a high ionisation, 'CIV-only' system. Maybe due to 
this shortage of observational data there are several theoretical
models which try to explain the origin of this type of absorption.    
Stengler-Larrea et al. (1996) have determined that strong CIV systems, 
with column density
greater than log$N > 14.2 \pm 0.2 $ also present low ionisation CII 
associated absorption. This favours a simple model for the absorber galaxies 
very similar to the one described in Steidel (1993), where the difference 
between low and high ionization systems basically relates to impact parameter.
Stengler-Larrea et al. (1996) estimate that the outer halo responsible 
for the high-ionisation absorption has a radius $R_{out}\sim 1.4R_{in}$, 
where $R_{in}$ is the radius of the inner low-ionisation sphere. 
From the Steidel et al. (1994) law, if object S were at z=1.82, it would have 
a low-ionisation absorption radius of $\approx 8\arcsec$ and thus a 
maximal high-ionization halo radius of $\approx 11\arcsec$, which is
consistent with the QSO spectrum presenting CIV but not low-ionisation 
absorptions at z=1.82 (see Fig. 4). This would be the first detection of 
a high-ionisation absorber and could help to clarify the controversy 
about their origin.

Arag\'on-Salamanca et al. (1994) have imaged fields around CIV 
absorption systems at redshift $z\sim 1.6$ in the near-IR, 
and claim the detection of several candidates for the absorber galaxies
at distances $r < 6\arcsec$. They would not have considered object S 
as a candidate because of its rather large impact parameter. However,
they have selected multiple CIV systems, which seem to be the 
high redshift counterparts of the relatively well known MgII systems
as they nearly always present low ionisation associated absorption. 
In fact, all the CIV systems of Arag\'on-Salamanca et al. (1994) present 
MgII or CII absorption when the corresponding part of the spectrum is 
observed, which explains why they have smaller impact parameters.

 Another possibility is that object S is a companion to the DLA system. 
However, if we place object S at $z=2.467$, it would become uncomfortably 
bright, for nearly all the possible combinations of galaxy types and 
cosmologies. 

\titleb{Other objects}

A close examination of Fig 3a reveals the existence of some faint 
objects close to the QSO. These objects are not seen in the $K_S$ band 
image, and they are not detected at a meaningful signal-to-noise in the I 
band, so we assume that most of them are probably noise, excepting maybe
the wisp to the SE of the QSO, which can also be seen in Fig. 4 of 
Lowenthal et al. 1995, a frame obtained with a narrow band 
filter tuned to [OII]$\lambda$ 3727 at z=0.788. 

It is noteworthy (see Fig 1) that objects 20 and 22, 
which are the close pair of objects at the upper left of Fig 3, 
and object 8 which is at $\approx 50 \arcsec$ from the QSO present 
$I-K$ colours similar to object S and could be members of a group at 
its redshift. Objects 7 and 6 are also remarkable: they are bright, 
quite compact but non stellar and have the colours expected for the 
red envelope at $z=0.788$, so they may be companions to object C, 
from which they are at a distance of $\approx 50 \arcsec$. 
In any case, spectroscopy, and deeper and wider multiband imaging 
are clearly needed in order to firmly establish the possible existence 
of groups at these redshifts.

\titlea{Conclusions}

We have performed deep near IR and optical imaging of the field around 
Q0836+113 with the aim of detecting the absorption systems present 
in its spectra and obtained the following results:

\noindent -- The images reveal the existence of a red object 
$\approx 11 \arcsec$ south from the QSO, which could be the 
CIV absorber at $z=1.82$. If this is confirmed, it would be 
the first detection of a CIV system which has no low ionization 
absorption and would contribute to clarify the controversy regarding 
the origin of these absorptions.

\noindent -- After carefully subtracting the QSO, 
we do not detect, up to a $3\sigma$ limiting magnitude of $K_S=20.8$ the 
galaxy responsible for the DLA absorption system which has 
been claimed to be detected as a Ly$\alpha$ emitter at $z=2.467$

\noindent -- We propose that object SW may be the galaxy causing the
MgII absorption at redshift $z=0.368$  on the basis of 
the Steidel et al. (1994) relationship between absorption radius and 
$K$ absolute magnitude for intermediate redshift MgII absorbers. 

\noindent -- There are several red objects in the field which could be 
companions to the CIV absorber and to the 
$z=0.788$ MgII absorber.

\acknow{
The 3.5m telescope is operated by the Max-Planck-Institut f\"ur Astronomie
at the Centro Astron\'omico Hispano-Alem\'an in Calar Alto 
(Almer\'\i a, Spain). 
The Nordic Optical Telescope is operated on the island of La Palma 
by the Nordic Optical Telescope Scientific Association in the Spanish 
Observatorio del Roque de Los Muchachos of the Instituto de Astrof\'\i sica
 de Canarias. We thank Chuck Steidel, Mark Dickinson, Max Pettini and 
Ignacio Ferreras for their help and for useful comments. 
NB, EMG and JLS acknowledge financial 
support from the Spanish DGICYT, 
project PB92-0741. NB acknowledges a Spanish M.E.C. Ph.D. scholarship.  
}
\begref{References}

\ref Arag\'on-Salamanca, A., Ellis, R., Couch, W.J., Carter, D. 1993, 
MNRAS, 262, 764

\ref Arag\'on-Salamanca, A., Ellis, R., Schwartzenberg, J.-M.,  Bergeron, 
J.A. 1994, ApJ, 421, 27

\ref Bergeron, J.,  Boisse, P. 1991, A\&A, 243, 344 

\ref Bruzual, G.A., Charlot, S., 1993, ApJ, 405, 538 

\ref Caulet, A. 1991, quoted in Wolfe et al. 1992 as 'private communication'.

\ref Draper, P.W., Eaton, N. 1992, Starlink Project User Note 109.5, 
Rutherford Appleton Laboratory

\ref  Dickinson, M. 1995, To appear in Fresh Views on Elliptical Galaxies, 
ASP Conference Series, eds. A. Buzzoni, A.Renzini and A. Serrano

\ref Elias, J.H., Frogel, J.A., Mathews, K., Neugebauer, G. 1982, 
AJ, 87, 1029

\ref Francis, P.J., Woodgate, B.E., Warren, S.J. et al. 1996,
 to appear in ApJ

\ref Hunstead, R.W., Pettini, M., Fletcher, A. B. 1990, ApJ, 356, 23

\ref Hu, E.M., Songaila, A., Cowie, L.L., Hodapp, K.-W., 1993 ApJLett, 419, L13

\ref Lanzetta, K.M., Bowen, D.B., Tytler, D., Webb, J.K. 1995, ApJ, 442, 538

\ref Lowenthal, J.D., Hogan, C.J., Green, R.F. et al. 1995, ApJ, 451, 484

\ref Neugebauer, G., Green, R.F., Matthews, K., 
Schmidt, M., Soifer, B.T., Bennet, J. 1987, ApJS, 63, 615

\ref Pettini, M., King, D.L., Smith, L.J., Hunstead, R.W. 1995, 
in QSO Absorption Lines, proc. of the ESO Workshop held at Garching, 
Germany, ed. G. Meylan, (Springer-Verlag), 55 

\ref Steidel, C.C. 1993, in The Environment and Evolution of Galaxies, 
proc. of the 3rd Teton Astronomy Conference, eds. J.M. Shull 
and H.A. Thronson, (Dordrecht: Kluwer), 263

\ref Steidel, C.C. 1995, in QSO Absorption Lines, proc. of the ESO 
Workshop held at Garching, Germany, ed. G. Meylan, (Springer-Verlag), 139

\ref Steidel, C.C., Dickinson, M. and Persson, S.E.  1994, 
ApJLett, 437, L75

\ref Stengler-Larrea, E.A., Boksenberg, A., Gonz\'alez-Solares, 
E.A. 1996, to appear in MNRAS

\ref Turnshek, D.A., Wolfe, A.M., Lanzetta, K.M. et al. 1989, ApJ, 344, 567 

\ref Wolfe, A.M., Turnshek, D.A., Lanzetta, K.M., Oke, J.B. 1992, ApJ, 385, 151

\endref
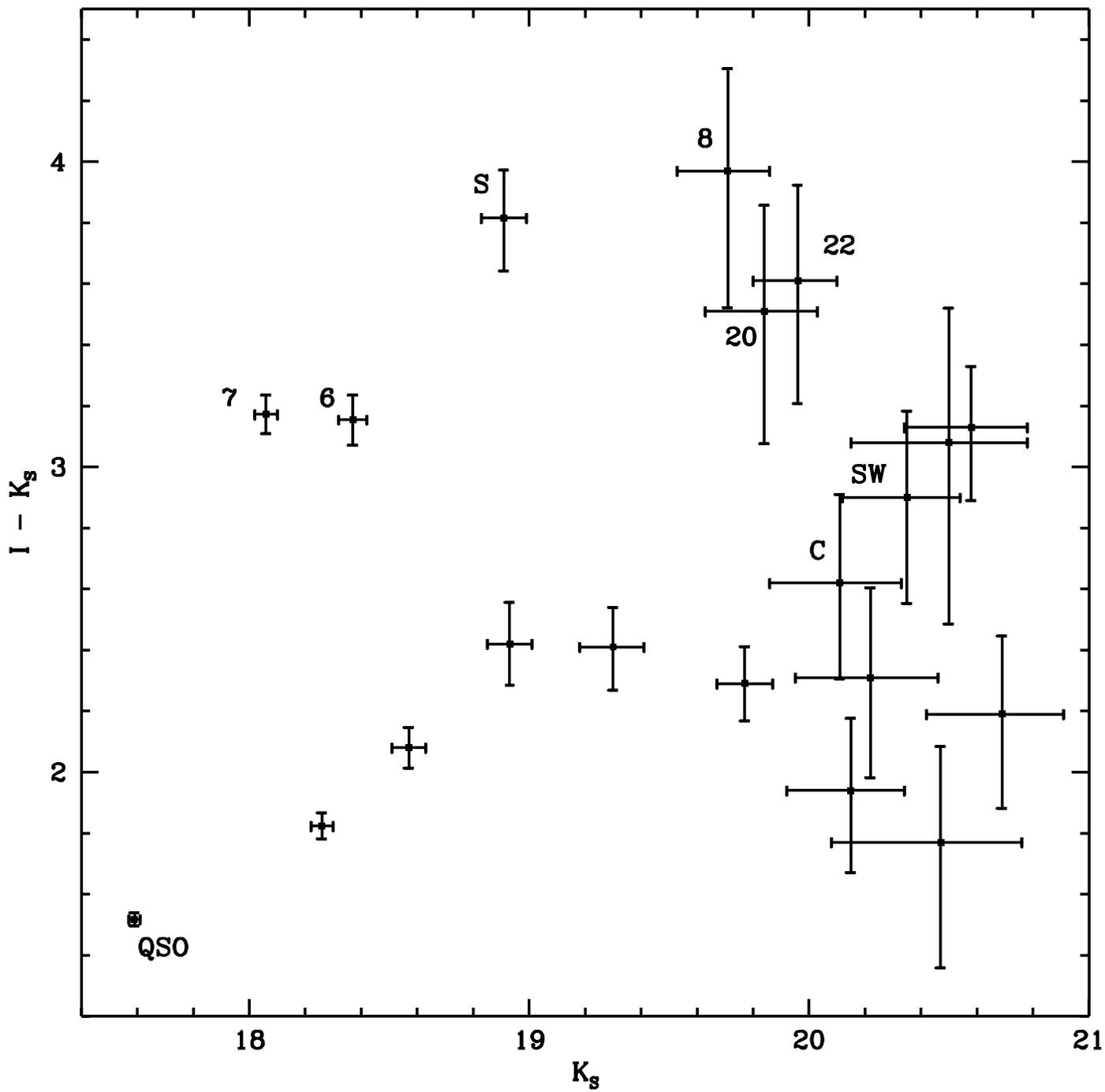
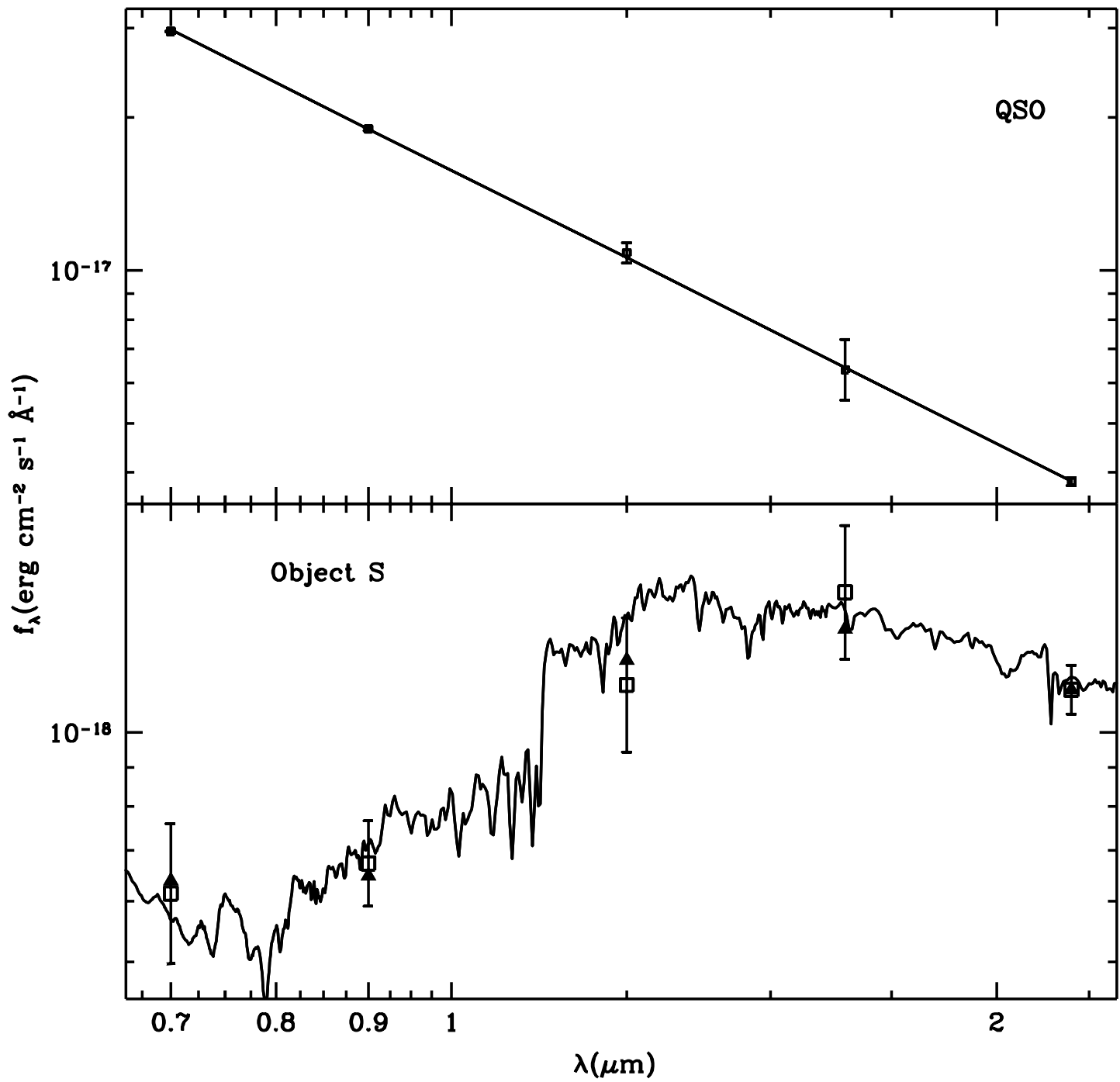
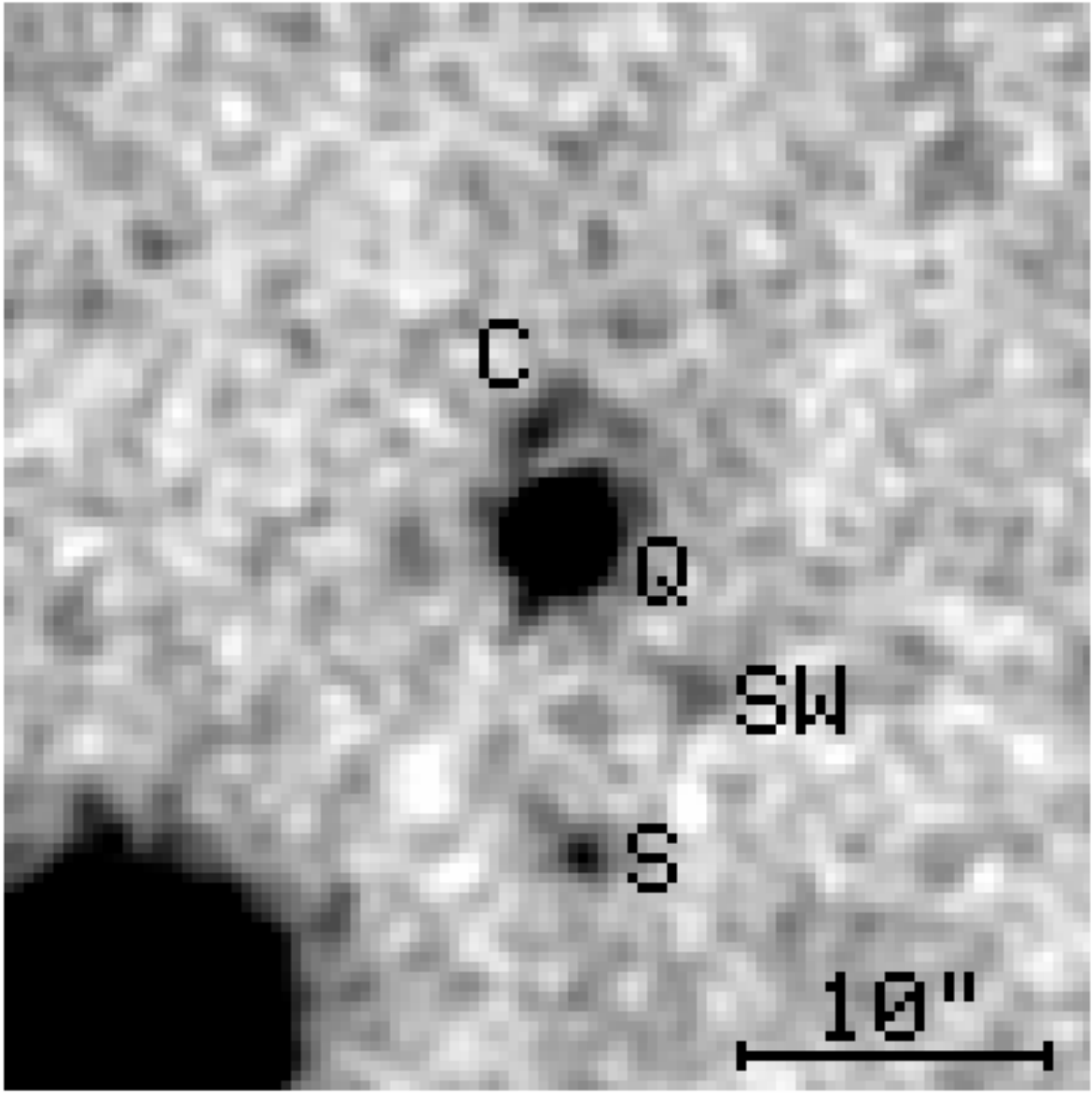
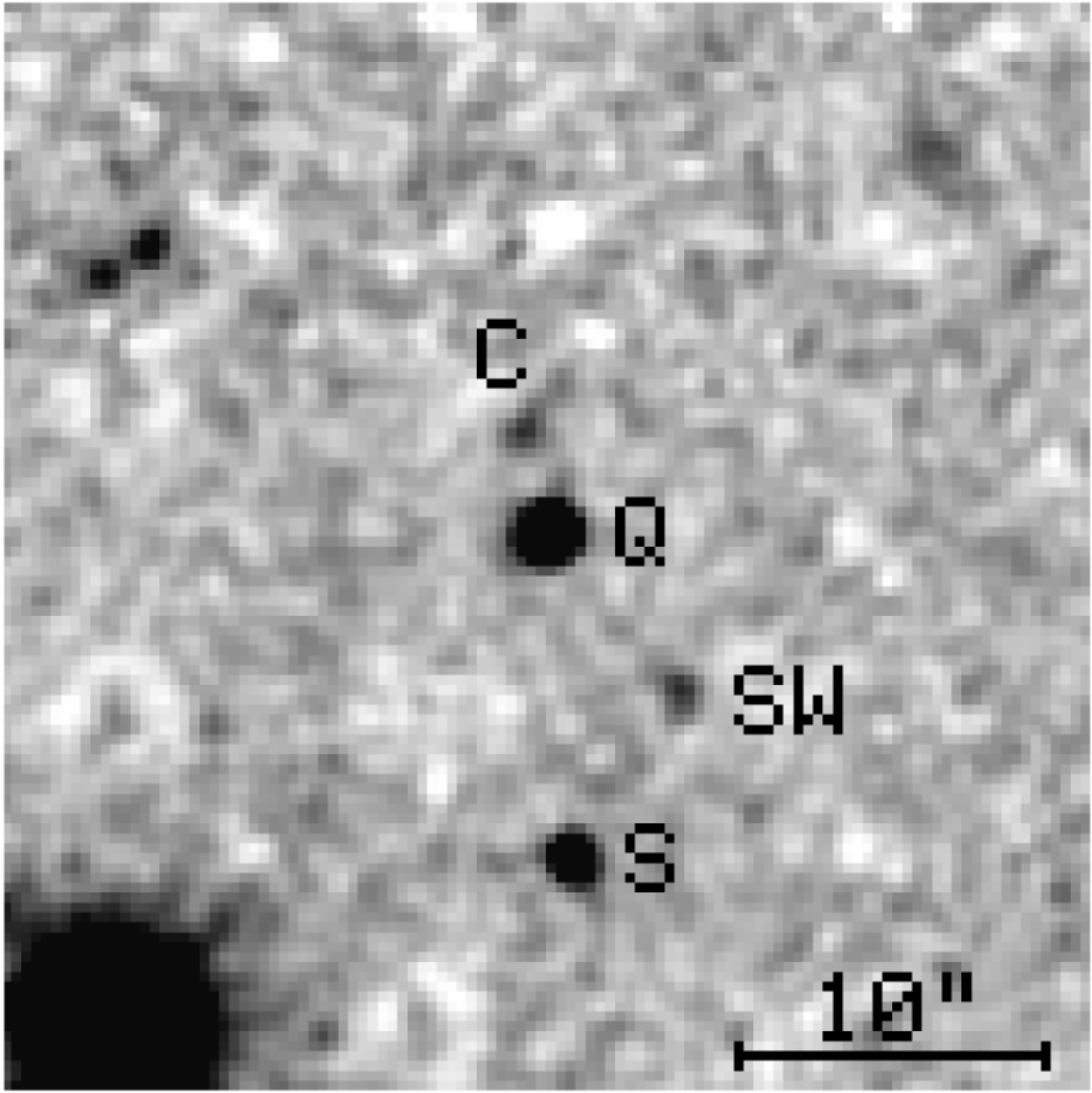
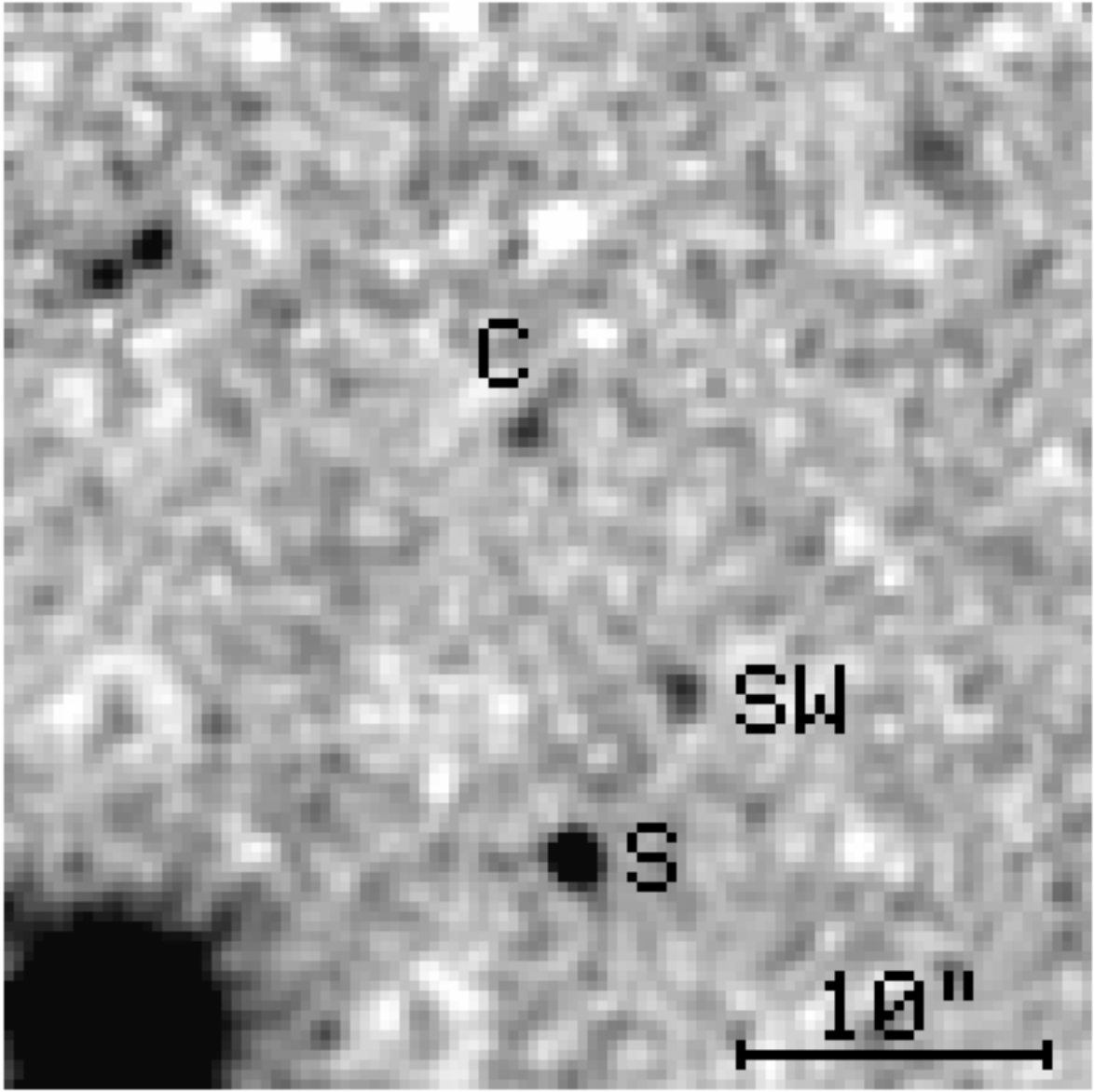
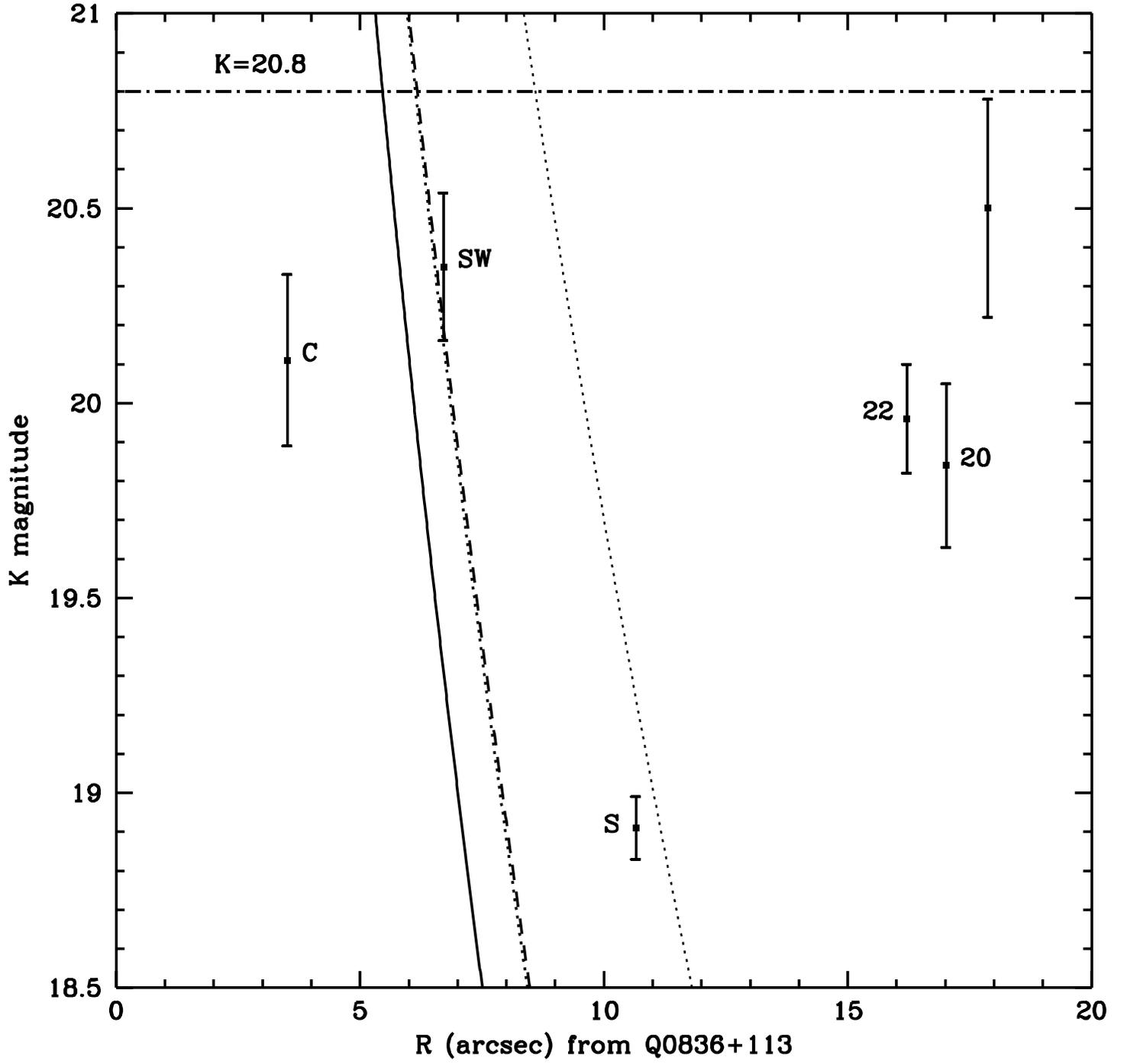
\bye